\documentclass{aa}

\usepackage{graphics}

\begin{document}

\thesaurus{03(03.20.1;13.25.3)}

\title{Spatial Corrections of ROSAT HRI Observations}
\author{D.E. Harris\inst{1}, J.D. Silverman\inst{1},
         G. Hasinger\inst{2} \and I. Lehmann\inst{2}}
\institute{Harvard-Smithsonian Center for Astrophysics, 60 Garden Street, Cambridge, MA 02138
    \and Astrophysikalisches Institut Potsdam, An der Sternwarte 16, D-14482 Potsdam, Germany}

%\altaffiltext{1}{present address: Center for Astrophysics, 60 Garden Street, Cambridge, MA 02138}

\date{Received date / Accepted date}
\offprints{D.E. Harris; harris@cfa.harvard.edu}

\titlerunning{Spatial Corrections of ROSAT HRI Data}
\authorrunning{Harris et al.}
\maketitle

\begin{abstract}

X-ray observations with the ROSAT High Resolution Imager (HRI) often
have spatial smearing on the order of 10$\arcsec$ (Morse 1994).  This
degradation of the intrinsic resolution of the instrument (5$\arcsec$)
can be attributed to errors in the aspect solution associated with the
wobble of the space craft or with the reacquisition of the guide
stars.  We have developed a set of IRAF/PROS and MIDAS/EXSAS routines
to minimize these effects.  Our procedure attempts to isolate aspect
errors that are repeated through each cycle of the wobble.  The method
assigns a 'wobble phase' to each event based on the 402 second period
of the ROSAT wobble.  The observation is grouped into a number of
phase bins and a centroid is calculated for each sub-image.  The
corrected HRI event list is reconstructed by adding the sub-images
which have been shifted to a common source position. This method has
shown $\sim$30\% reduction of the full width half maximum (FWHM) of an
X-ray observation of the radio galaxy 3C 120.  Additional examples are
presented.

\end{abstract}


\section{Introduction}

Spatial analysis of ROSAT HRI observations is often plagued by poor
aspect solutions, precluding the attainment of the potential
resolution of about 5''.  In many cases (but not all), the major
contributions to the degradation in the effective Point Response
Function (PRF) come from aspect errors associated either with the
ROSAT wobble or with the reacquisition of the guide stars.

To avoid the possibility of blocking sources by the window support
structures (Positional Sensitive Proportional Counter) or to minimize
the chance that the pores near the center of the microchannel plate
would become burned out from excessive use (High Resolution Imager),
the satellite normally operates with a constant dither for pointed
observations.  The period of the dither is 402s and the phase is tied
to the spacecraft clock.  Any given point on the sky will track back
and forth on the detector, tracing out a line of length $\approx$~3
arcmin with position angle of 135$^{\circ}$ in raw detector
coordinates (for the HRI).  Imperfections in the star tracker (see
section~\ref{sec:MM}) can produce an erroneous image if the aspect
solution is a function of the wobble track on the CCD of the star
tracker.

This work is similar to an analysis by Morse (1994) except that we do
not rely on a direct correlation between spatial detector coordinates
and phase of the wobble.  Moreover, our method addresses the
reacquisition problem which produces the so-called cases of
``displaced OBIs''.  An ``OBI'' is an observation interval, normally
lasting for 1 ks to 2 ks (i.e.  a portion of an orbit of the
satellite).  A new acquisition of the guide stars occurs at the
beginning of each OBI and we have found that different aspect
solutions often result.  Occasionally a multi-OBI observation consists
of two discrete aspect solutions.  A recent example (see
section~\ref{sec:120B}) showed one OBI for which the source was
10$^{\prime\prime}$ north of its position in the other 17 OBIs.  Note
that this sort of error is quite distinct from the wobble error.

Throughout this discussion, we use the term ``PRF'' in the dynamic
sense: it is the point response function realized in any given
situation: i.e.  that which includes whatever aspect errors are
present.  We start with an observation for which the PRF is much worse
than it should be.  We seek to improve the PRF by isolating the
offending contributions and correcting them if possible or rejecting
them if necessary.

\section{Model and Method}\label{sec:MM}

The ``model'' for the wobble error assumes that the star tracker's CCD has
some pixels with different gain than others.  As the wobble moves the
de-focused star image across the CCD, the centroiding of the stellar
image gets the wrong value because it is based on the relative response
from several pixels.  If the roll angle is stable, it is likely that the
error is repeated during each cycle of the wobble since the star's path
is over the same pixels (to a first approximation if the aspect `jitter'
is small compared to the pixel size of $\approx$~1 arcmin).  What is not
addressed is the error in roll angle induced by erroneous star
positions.  If this error is significant, the centroiding technique with
one strong source will fix only that source and its immediate environs. 

The correction method assigns a 'wobble phase' to each event; then
divides each OBI (or other suitably defined time interval) into a number
of wobble phase bins.  The centroid of the reference source is measured
for each phase bin.  The data are then recombined after applying x and y
offsets in order to ensure that the reference source is aligned for each
phase bin.  What is required is that there are enough counts in the
reference source to obtain a reliable centroid.  Variations of this
method for sources weaker than approx 0.1 count/s involve using all
OBIs together before dividing into phase bins.  This is a valid approach
so long as the nominal roll angle is stable (i.e.  within a few tenths
of a degree) for all OBIs, and so long as major shifts in the aspect
solutions of different OBIs are not present.

\section{Diagnostics}

Our normal procedure for evaluation is to measure the FWHM (both the
major and minor axes) of the observed response on a map smoothed with
a 3$^{\prime\prime}$ Gaussian.  For the best data, we find the
resulting FWHM is close to 5.7$^{\prime\prime}$.  While there are many
measures of source smearing, we prefer this approach over measuring
radial profiles because there is no uncertainty relating to the
position of the source center; we are normally dealing with elliptical
rather than circular distributions; and visual inspection of the two
dimensional image serves as a check on severe abnormalities.
It has been our experience that when we are able to reduce the
FWHM of the PRF, the wings of the PRF are also reduced.

\begin{figure}
  \resizebox{\hsize}{!}{\includegraphics{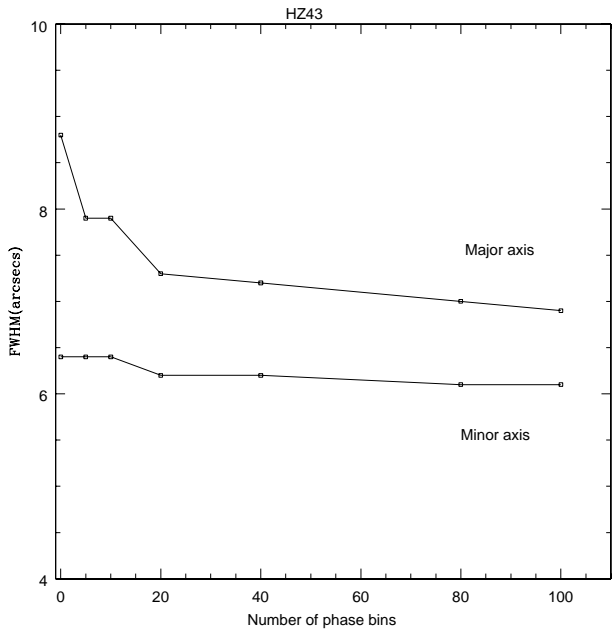}}
 
  \caption{The FWHM of a HZ43 (observation number rh142545)
observation was measured for multiple dewobble runs while increasing
the number of phase bins.}
 
  \label{fig:hz43}
\end{figure}

\subsection{Wobble Errors}

If the effective PRF is evaluated for each OBI separately, the wobble
problem is manifest by a degraded PRF in one or more OBIs.  Most OBIs
contain only the initial acquisition of the guide stars, so when the
PRF of a particular OBI is smeared, it is likely to be caused by the
wobble error and the solution is to perform the phased `de-wobbling'.

\subsection{Misplaced OBI}

For those cases where each OBI has a relatively good PRF but the
positions of each centroid have significant dispersion, the error
cannot be attributed to the wobble.  We use the term `misplaced OBI'
to describe the situation in which a different aspect solution is
found when the guide stars are reacquired.  In the worst case,
multiple aspect solutions can produce an image in which every source
in the field has a companion displaced by anywhere from 10 to 30
arcsec or more.  When the separation is less than 10 arcsec, the
source can appear to have a tear drop shape (see
section~\ref{sec:120A}) or an egg shape.  However, depending on the
number of different aspect solutions, almost any arbitrary distortion
to the (circularly symmetric) ideal PRF is possible.  The fix for
these cases is simply to find the centroid for each OBI, and shift
them before co-adding (e.g., see Morse et al. 1995).

\section{IRAF/PROS Implementation}

The ROSAT Science Data Center (RSDC) at SAO has developed scripts to
assist users in evaluating individual OBIs and performing the
operations required for de-wobbling and alignment.  The scripts are
available from our anonftp area: sao-ftp.harvard.edu.  cd to
pub/rosat/dewob.

An initial analysis needs to be performed to determine the stable roll
angle intervals, to check for any misalignment of OBIs and to examine
the guide star combinations.  These factors together with the source
intensity are important in deciding what can be done and the best
method to use.

\subsection{OBI by OBI Method}\label{sec:ObyO}

If the observation contains a strong source ($\ge$~0.1 counts/s) near
the field center (i.e. close enough to the center that the mirror
blurring is not important), then the preferred method is to dewobble
each OBI.  The data are thus divided into n~$\times$~p qpoe files (n =
number of OBIs; p = number of phase bins).  The position of the
centroid of the reference source is determined and each file is
shifted in x and y so as to align the centroids from all OBIs and all
phase bins.  The data are then co-added or stacked to realize the
final image (qpoe file).

\subsection{Stable Roll Angle Intervals}

For sources weaker than 0.1 counts/s, it is normally the case that there
are not enough counts for centroiding when 10 phase bins are used.
If it is determined that there are no noticeable shifts between OBIs,
then it is possible to use many OBIs together so long as the roll
angle does not change by a degree or more.

\subsection{Method for Visual Inspection}
 
On rare occasions, it may be useful to examine each phase bin visually
to evaluate the segments in order to decide if some should be deleted
before restacking for the final result.  We have found it useful to do
this via contour diagrams of the source.  This approach can be labor
intensive if there are a large number of OBIs and phase bins but
scripts we provide do most of the manipulations.

\section{MIDAS/EXSAS Implementation}
 
The X-ray group at the Astrophysical Institute Potsdam (AIP) has
developed some MIDAS/EXSAS routines to correct for the ROSAT wobble
effect.  The routines can be obtained by anonymous ftp from ftp.aip.de
at directory pub/users/rra/wobble. The correction procedure works
interactively in five main steps:
 
\begin{itemize}
\item{Choosing of a constant roll angle interval}
\item{Folding the data over the 402 sec wobble period}
\item{Creation of images using 5 or 10 phase intervals}
\item{Determining the centroid for the phase resolved images}
\item{Shifting the photon X/Y positions in the events table}
\end{itemize}
 
We have tested the wobble correction procedures for 21 stars and 24
galaxies of the ROSAT Bright Survey using archival HRI data. The
procedures work successfully down to an HRI source count rate of about
0.1 counts/s.  In the case of lower count rates the determination of
the centroid position failed because of the few photons available in
the phase-binned images.  The number of phase bins which can be used
is of course dependent on the X-ray brightness of the source.

\section{Limitations}

We briefly describe the effects which limit the general use of the
method.  In so doing, we also indicate the process one can use in
deciding if there is a problem, and estimating the chances of
substantial improvement.

\subsection{Presence of Aspect Smearing}

The FWHM of all sources in the field should
be~$\ge$~7$^{\prime\prime}$~(after smoothing with a 3$^{\prime\prime}$
Gaussian).  If any source is smaller than this value, it is likely
that aspect problems are minimal and little is to be gained by
applying the dewobbling method.

If there is only a single source in the field, without {\it a~priori}
knowledge or further analysis it is difficult to determine whether a
distribution significantly larger than the ideal PRF is caused by
source structure or aspect smearing.  The best approach in this case
is to examine the image for each OBI separately to see if some or all
are smaller than the total image (i.e. OBI aspect solutions are
different).

\subsection{Wobble Phase}

It is important that the phase of the wobble is maintained.  This is
ensured if there is no 'reset' of the space craft clock during an
observation.  If an observation has a begin and end time/date that
includes a reset, it will be necessary to divide the data into two
segments with a time filter before proceeding to the main analysis.
Dates of clock resets (Table 1) are provided by MPE:
http://www.ROSAT.mpe-garching.mpg.de/$\sim$prp/timcor.html.

\begin{tabular*}{45mm}{cl}
\multicolumn{2}{c}{Table 1} \\
\multicolumn{2}{c}{ROSAT Clock Resets}\\
\hline\\
Year &Day \\
\hline
90 & 151.87975 (launch)\\
91 & 25.386331\\
92 & 42.353305\\
93 & 18.705978\\
94 & 19.631352\\
95 & 18.169322\\
96 & 28.489871\\
97 & 16.069990\\
98 & 19.445738\\
\end{tabular*}
\\
\\

\begin{figure}[h]
\begin{minipage}{8.5cm}
 \rotatebox{-90}{
  \resizebox{7.0cm}{!}{\includegraphics{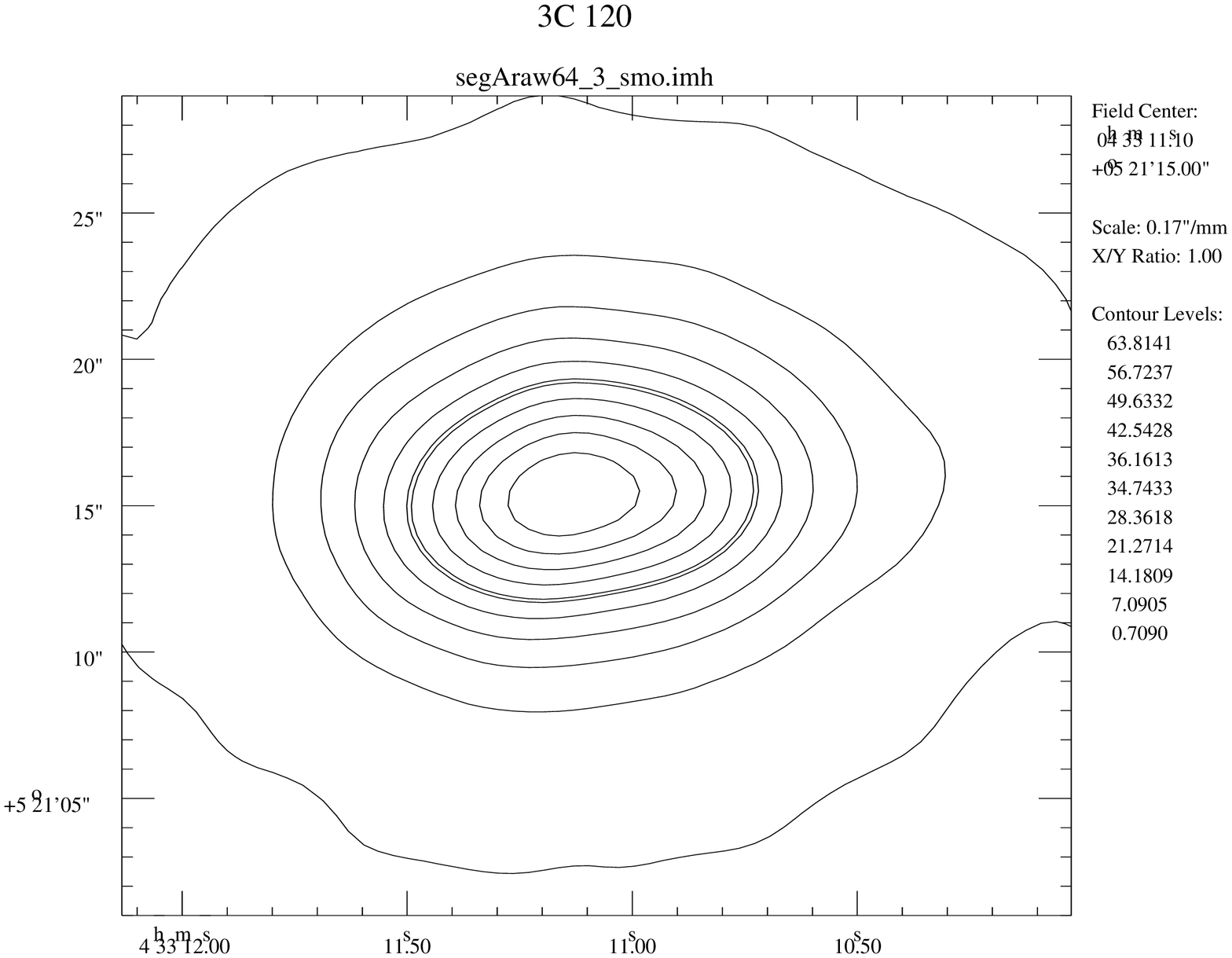}}} 
\caption{The original data for 3C 120 (segment A, rh702080n00),
  smoothed with a Gaussian of FWHM = 3$^{\prime\prime}$.  The peak
  value on the map is 70.9 counts per 0.5$^{\prime\prime}$ pixel.
  Contour levels are 1, 10, 20, 30, ...  90\% of the peak value, with
  the 50\% contour, doubled.  The nominal roll angle is -167$^{\circ}$
  and the wobble direction is at PA = 122$^{\circ}$. The FWHM of this
  smoothed image is
  11.6$^{\prime\prime}~\times$~7.4$^{\prime\prime}$.}
  \label{fig:120A}
%\end{figure}

%\begin{figure}
  \rotatebox{-90}{
  \resizebox{7.0cm}{!}{\includegraphics{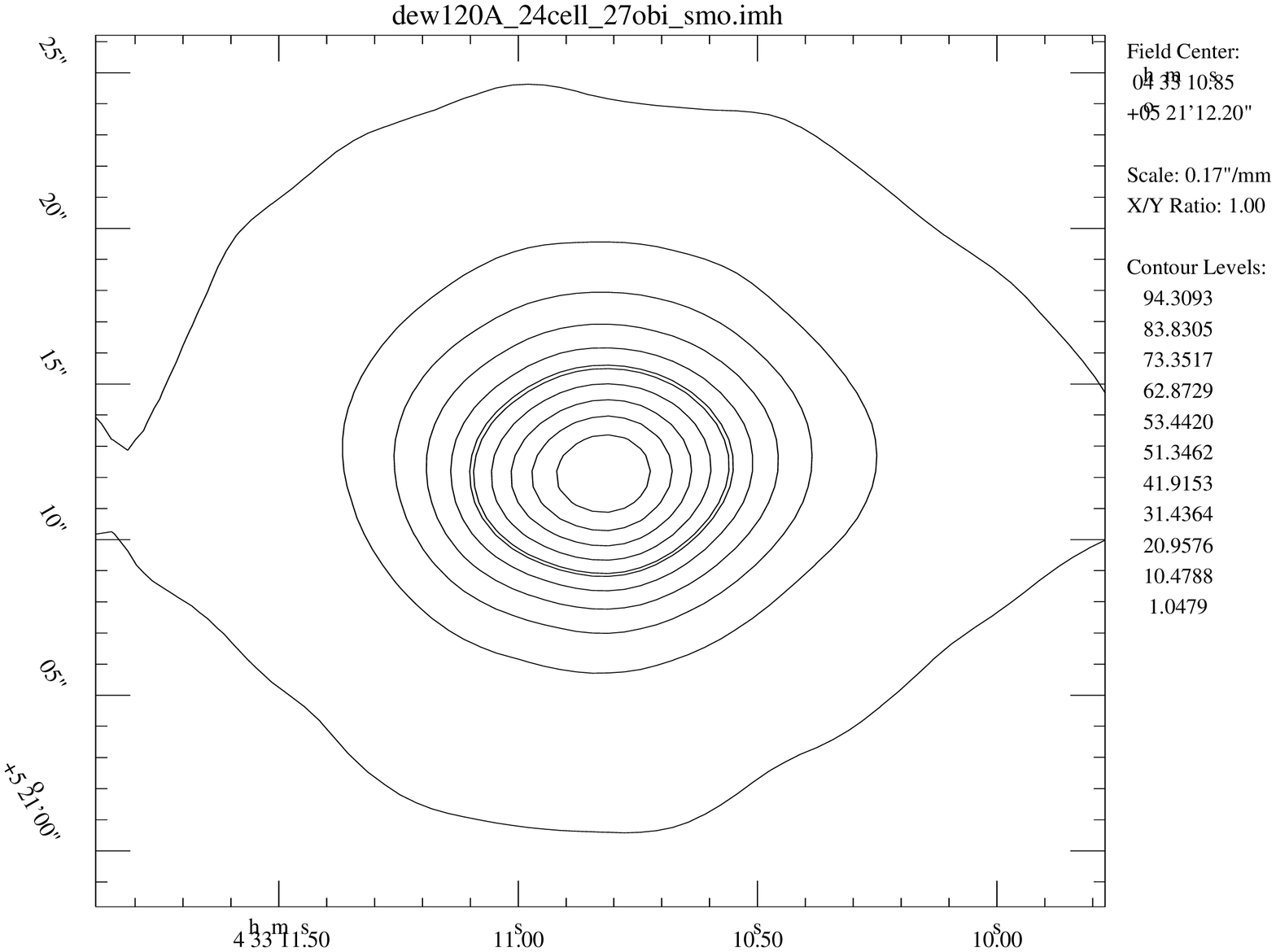}}} 
\caption{The results after dewobbling 3C 120A, smoothed with a
  Gaussian of FWHM = 3$^{\prime\prime}$.  The peak value on the map is
  now 104.8 counts per 0.5$^{\prime\prime}$ pixel.  Contour levels are
  1, 10, 20, 30, ... 90\% of the peak value, with the 50\% contour,
  doubled.  The FWHM of this smoothed image
  is 8.1$^{\prime\prime}~\times$~6.7$^{\prime\prime}$.}
  \label{fig:120Ade}
\end{minipage}
\end{figure}

\subsection{Characteristics of the Reference Source}

In most cases, the reference source (i.e. the source used for
centroiding) will be the same as the target source, but this is not
required.  Ideally, the reference source should be unresolved in the
absence of aspect errors and it should not be embedded in high
brightness diffuse emission (e.g. the core of M87 does not work
because of the bright emission from the Virgo Cluster gas).  Both of
these considerations are important for the operation of the
centroiding algorithm, but neither is an absolute imperative.  For
accurate centroiding, the reference source needs to stand
well above any extended component.

Obviously the prime concern is that there be enough counts in a phase
bin to successfully measure the centroid.  The last item is usually
the determining factor, and as a rule of thumb, it is possible to use
10 phase bins on a source of 0.1 counts/s.  We have tested a strong
source to see the effect of increasing the number of phase bins.  In
Fig.~\ref{fig:hz43}, we show the results of several runs on an
observation of HZ 43 (12 counts/s).  This figure demonstrates that ten
phase bins is a reasonable choice, but that there is little to be
gained by using more than 20 phase bins.

\section{Examples}

\subsection{3C 120}

3C 120 is a nearby radio galaxy (z=0.033) with a prominent radio jet
leaving the core at PA $\approx 270^{\circ}$.  The ROSAT HRI
observation was obtained in two segments, each of which had aspect
problems.  Since the average source count rate is 0.8 count/s, the
X-ray emission is known to be highly variable (and therefore most of
its flux must be unresolved), and each segment consisted of many OBIs,
we used these observations for testing the dewobbling scripts.

\subsubsection{Segment A: Two aspect solutions, both found multiple
times}\label{sec:120A}
 
 The smoothed data (Figure~\ref{fig:120A}) indicated that in addition
to the X-ray core, a second component was present, perhaps associated
with the bright radio knot 4$^{\prime\prime}$ west of the core.  When
analyzing these two components for variability, it was demonstrated
that most of the emission was unresolved, but that the aspect solution
had at least two different solutions, and that the change from one to
the other usually coincided with OBI boundaries.  The guide star
configuration table showed that a reacquisition coincided with the
change of solution.

The 24 OBIs comprising the 36.5 ksec exposure were obtained between
96Aug16 and 96Sep12.  Because 3C 120 is close to the ecliptic, the
roll angle hardly changed, and our first attempts at dewobbling
divided the data into 2 'stable roll angle intervals'.  This effort
made no noticeable improvement.

We then used the method described in section \ref{sec:ObyO}.  The
results are shown in Figure~\ref{fig:120Ade}.  It can be seen that a
marked improvement has occurred, but some of the E-W smearing remains.

\begin{figure}[h]
\begin{minipage}{8.5cm}
  \rotatebox{-90}{
  \resizebox{7.0cm}{!}{\includegraphics{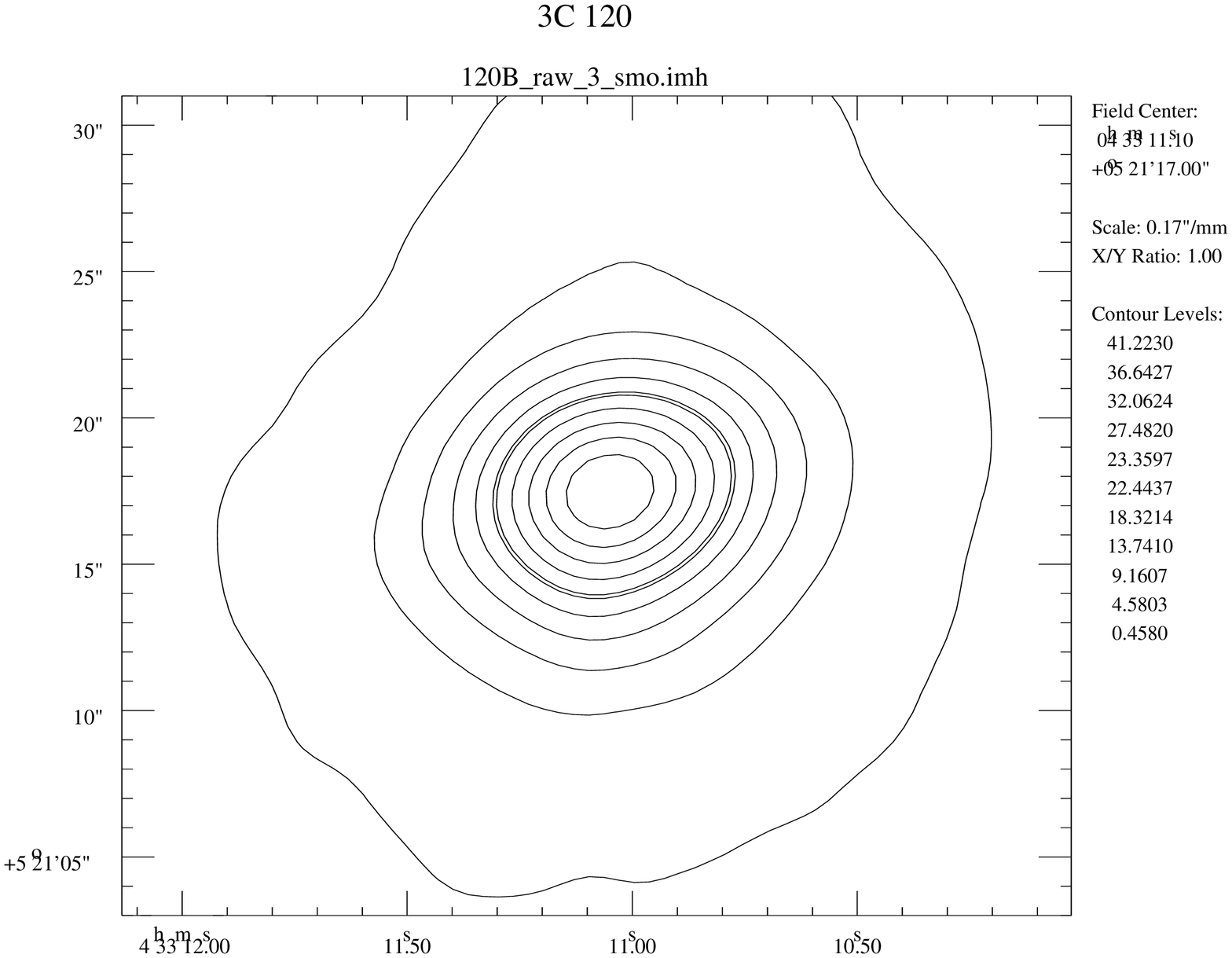}}} 
\caption{The original data of 3C 120 (segment B, rh702080a01),
  smoothed with a Gaussian of FWHM = 3$^{\prime\prime}$.  The peak
  value on the map is 45.8 counts per 0.5$^{\prime\prime}$ pixel.  The
  contour levels are the same percentage values as those of
  Fig.~\ref{fig:120A}.  The roll angle is 8$^{\circ}$ and the wobble
  PA is 127$^{\circ}$.  FWHM for this image is
  8.0$^{\prime\prime}~\times$~6.7$^{\prime\prime}$.}  \label{fig:120B}
%\end{figure}
 
%\begin{figure}
  \rotatebox{-90}{
  \resizebox{7.0cm}{!}{\includegraphics{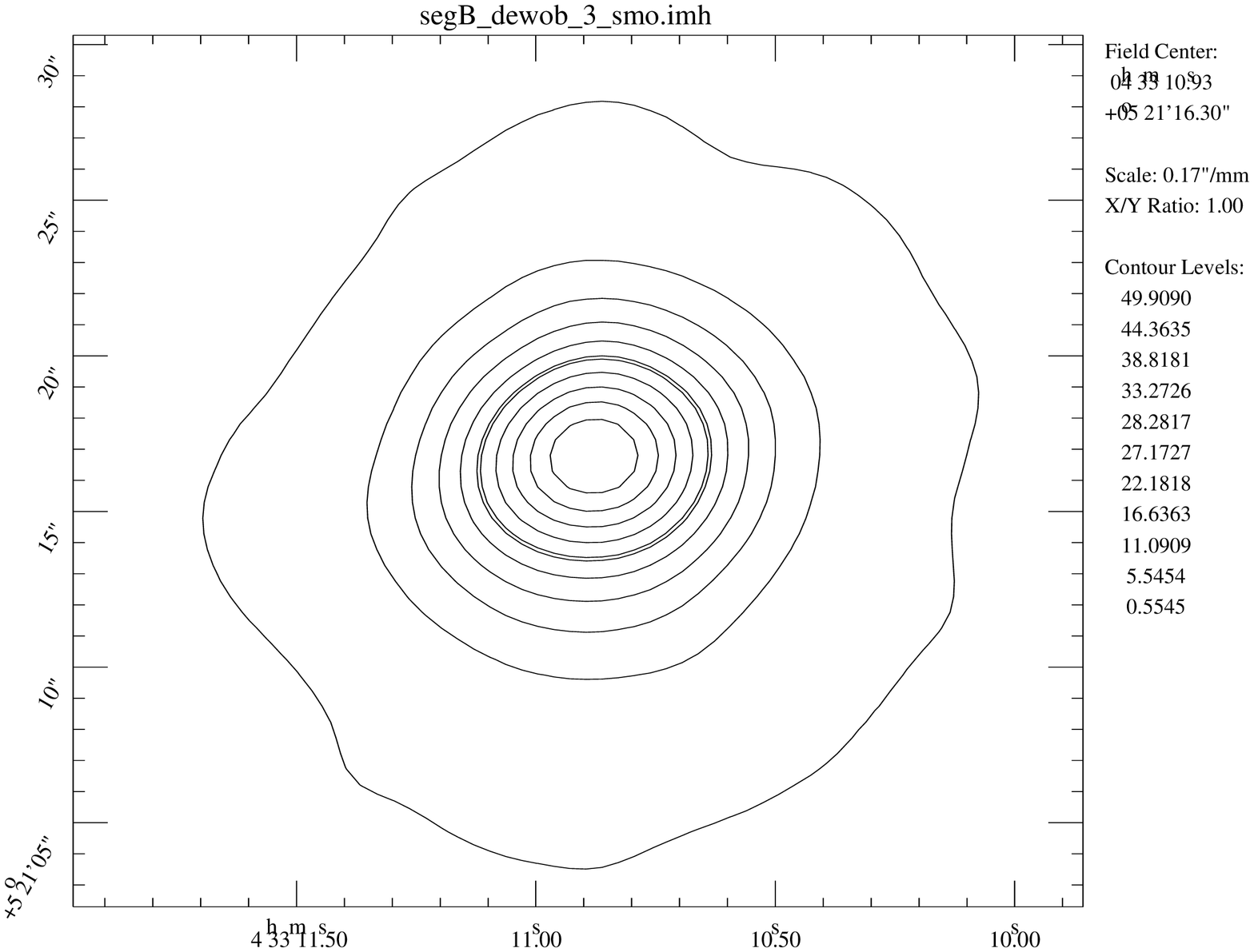}}} \caption{The
  results of 3C 120 (segment B) after dewobbling.  The contour levels
  are the same percentage values as those of Fig.~\ref{fig:120B}, but
  the peak is now 55.4.  The FWHM is
  7.2$^{\prime\prime}~\times$~6.5$^{\prime\prime}$.} \label{fig:120Bde}

\end{minipage}
\end{figure}

\subsubsection{Segment B: A single displaced OBI}\label{sec:120B}

The second segment of the 3C 120 observation was obtained in 1997 March.
In this case, only one OBI out of 17 was displaced.  It was positioned
10$^{\prime\prime}$ to the north of the other positions, producing a
low level extension (see Fig.~\ref{fig:120B}).  After dewobbling, that
feature is gone, the half power size is reduced, and the peak value is
larger (Fig.~\ref{fig:120Bde}).

\subsection{M81}

\begin{figure}[h]
\begin{minipage}{8.5cm}
  \resizebox{\hsize}{!}{\includegraphics{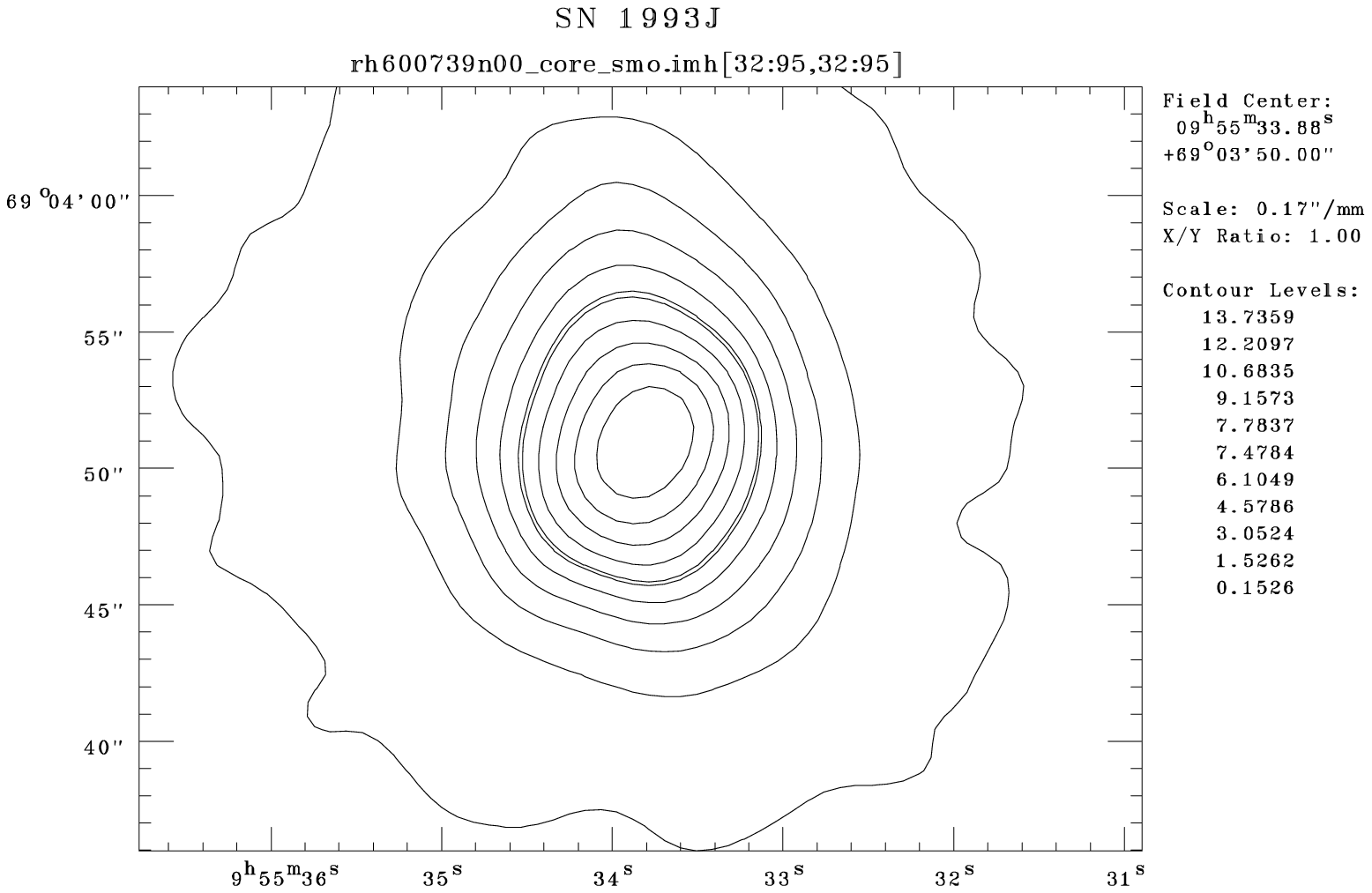}} \caption{The
  original M81 data (rh600739), smoothed with a Gaussian of FWHM =
  3$^{\prime\prime}$.  The peak value on the map is 15.3 counts per
  0.5$^{\prime\prime}$ pixel.  The contour levels are 1, 10, 20, 30,
  40, 50 (the 50\% contour, doubled), 60, 70, 80, and 90 percent of
  the peak value.  The nominal roll angle is 135$^{\circ}$ and the
  wobble direction is 0$^{\circ}$. The
  FWHM of this smoothed image is
  10.4$^{\prime\prime}~\times$~7.5$^{\prime\prime}$.}
  \label{fig:m81A}
%\end{figure}
 
%\begin{figure}
  \resizebox{\hsize}{!}{\includegraphics{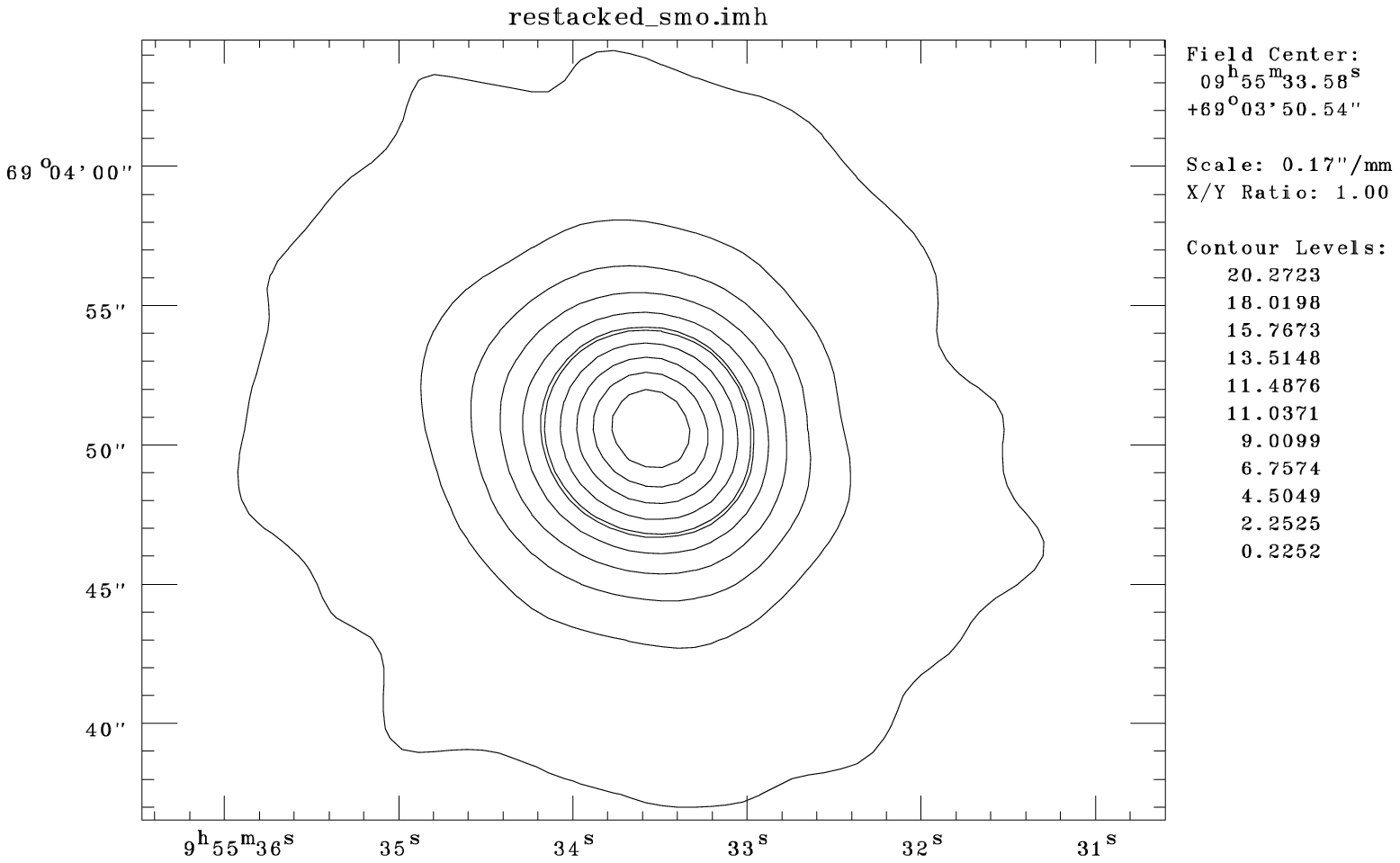}}
  \caption{The results after dewobbling of M81 smoothed with a
Gaussian of FWHM = 3$^{\prime\prime}$.  The peak value on the map is
22.5 counts per 0.5$^{\prime\prime}$ pixel.  The contour levels are
1, 10, 20, 30, 40, 50 (the 50\% contour, doubled),
60, 70, 80, and 90 percent of the peak value. Ten phase bins have been
  used.  The
  FWHM of this smoothed image is
  7.2$^{\prime\prime}~\times$~6.5$^{\prime\prime}$.}
  \label{fig:m81B}
\end{minipage}
\end{figure}

M81 is dominated by an unresolved nuclear source.  The count rate is
0.31 count/s.  The observation has 14 OBIs for a total exposure of
19.9 ks.  Figure~\ref{fig:m81A} shows the data from SASS processing.
After running the `OBI by OBI' method, the source is more circularly
symmetric, has a higher peak value, and a smaller FWHM (Fig.~\ref{fig:m81B}).

\subsection{NGC 5548}

This source was observed from 25 June to 11 July 1995 for a livetime of
53 ks with 33 OBIs.  The average count rate was 0.75 counts/s and
the original data had a FWHM =
8.2$^{\prime\prime}\times$6.8$^{\prime\prime}$.  Most of the OBIs
appeared to have a normal PRF
but a few displayed high distortion.  After applying
the OBI by OBI method, the resulting FWHM was 6.3$^{\prime\prime}$ in
both directions and the peak value on the smoothed map increased from
138 to 183 counts per 0.5$^{\prime\prime}$ pixel.

\subsection{RZ Eri}

The observation of this star was reduced in MIDAS/EXSAS.  The source
has a count rate of 0.12 count/s.  The reduction selected only a group
of the OBIs which comprised a 'stable roll angle interval'; almost
half the data were rejected.  The original smoothed image had a FWHM =
8.4$^{\prime\prime}\times$6.6$^{\prime\prime}$.  After dewobbling, the
resulting FWHM was 6.9$^{\prime\prime}\times$5.8$^{\prime\prime}$.

\section{Summary}

We have developed a method of improving the spatial quality of ROSAT
HRI data which suffer from two sorts of aspect problems.  This
approach requires the presence of a source near the field
center which has a count rate of $\approx$ 0.1 counts/s or greater.
Although the method does not fix all bad aspect problems, it produces
marked improvements in many cases.

\section{Acknowledgments}
 
We thank M. Hardcastle (Bristol) for testing early versions of the
software and for suggesting useful improvements.  J. Morse contributed
helpful comments on the manuscript.  The 3C 120 data were kindly
provided by DEH, A. Sadun, M. Vestergaard, and J. Hjorth (a paper is
in preparation).  The other data were taken from the ROSAT archives.
The work at SAO was supported by NASA contract NAS5-30934.

\end{document}